\documentclass[a4paper]{article}

\usepackage{INTERSPEECH2015}

\usepackage{graphicx}
\usepackage{amssymb,amsmath,bm}
\usepackage{textcomp}
\usepackage{epstopdf,subfigure,tabularx,multirow}

\ninept

\title{Improving Short Utterance PLDA Speaker Verification using SUV Modelling and Utterance Partitioning Approach}

\makeatletter
\def\name#1{\gdef\@name{#1\\}}
\makeatother 
\name{{\em Ahilan Kanagasundaram$^{*+}$, David Dean$^{*}$, Sridha Sridharan$^{*}$ and Clinton Fookes$^{*}$}}
\address{Speech and Audio Research Laboratory$^{*}$  \\
	Queensland University of Technology, Brisbane, Australia$^{*}$ \\
	{\small \tt \{a.kanagasundaram, d.dean, s.sridharan, c.fookes\}@qut.edu.au}$^{*}$ \\
	Electrical \& Electronic Engineering, Faculty of Engineering$^{+}$ \\
	University of Jaffna, Jaffna, Sri Lanka$^{+}$\\
	{\small \tt ahilan@eng.jfn.ac.lk}$^{+}$
}

\begin{document}
%
\maketitle
\begin{abstract}
This paper analyses the short utterance probabilistic linear discriminant analysis~(PLDA) speaker verification with utterance partitioning and short utterance variance~(SUV) modelling approaches. Experimental studies have found that instead of using single long-utterance as enrolment data, if long enrolled-utterance is partitioned into multiple short utterances and average of short utterance i-vectors is used as enrolled data, that improves the Gaussian PLDA~(GPLDA) speaker verification. This is because short utterance i-vectors have speaker, session and utterance variations, and utterance-partitioning approach compensates the utterance variation. Subsequently, SUV-PLDA is also studied with utterance partitioning approach, and utterance-partitioning-based SUV-GPLDA system shows relative improvement of 9\% and 16\% in EER for NIST 2008 and NIST 2010 truncated 10sec-10sec evaluation condition as utterance-partitioning approach compensates the utterance variation and SUV modelling approach compensates the mismatch between full-length development data and short-length evaluation data.
\end{abstract}
\noindent{\bf Index Terms}: speaker verification, i-vectors, PLDA, SUV, utterance partitioning
\section{Introduction}
A significant amount of speech is required for speaker model enrolment and verification, especially in the presence of large intersession variability, which has limited the widespread use of speaker verification technology in everyday applications. Reducing the amount of speech required for development, training and testing while obtaining satisfactory performance has been the focus of a number of recent studies in state-of-the-art speaker verification design, including joint factor analysis~(JFA), i-vectors, probabilistic linear discriminant analysis~(PLDA) and support vector machines~(SVM)~\cite{Vogt2008a,Kanagasundaram2011,McLaren2010b,Kanagasundaram2012a,Kanagasundaram2014(Submitted),Kanagasundaram2013b}. Continuous research on this field has been ongoing to address the robustness of speaker verification technologies under such conditions.

Previous research studies had found that long utterance i-vectors contain two source of variation: changing speaker characteristics, and changing channel (or session) characteristics~\cite{Dehak2010}. Recently it was found that short utterance i-vectors vary due to speaker, session and linguistic content~(utterance variation)~\cite{Kenny2013,Kanagasundaram2014(Submitted)}. In typical PLDA speaker verification, a single utterance is used as enrolment data. In this paper, instead of using a single long-utterance as enrolment data, long-enrolment utterance is partition into multiple short-enrolment utterances and the average i-vector over the short utterances is used as enrolment data to improve the short utterance speaker verification system. Recently, we have also introduced short utterance variance~(SUV) modelling to PLDA speaker verification system to compensate the session and utterance variations~\cite{Kanagasundaram2013(tobesubmitted)}. Subsequently, in this paper, we also investigate the utterance-partitioning-based GPLDA speaker verification with SUV modelling approach.

This paper is structured as follows: Section~\ref{sec:i-vector feaure extraction} details the i-vector feature extraction techniques. Section~\ref{sec:SUV added i-vectors} details the short utterance variance added i-vector feature extraction approach. Section~\ref{sec:len-norm GPLDA} explains the GPLDA based speaker verification system. The experimental protocol and corresponding results are given in Section~\ref{sec:method} and Section~\ref{sec:results and discussions}. Section~\ref{sec:conclusion} concludes the paper.
\section{I-vector feature extraction} \label{sec:i-vector feaure extraction}
I-vectors represent the GMM super-vector by a single total-variability subspace. This single-subspace approach was motivated by the discovery that the channel space of JFA contains information that can be used to distinguish between speakers~\cite{Dehak2009a}. An i-vector speaker and channel dependent GMM super-vector can be represented by,
\begin{eqnarray}
    \boldsymbol{\mu} & = & \textbf{m} + \textbf{Tw},
\end{eqnarray}
where $\textbf{m}$ is the same universal background model~(UBM) super-vector used in the JFA approach and $\textbf{T}$ is a low rank total-variability matrix. The total-variability factors~($\textbf{w}$) are the i-vectors, and  are normally distributed with parameters $\emph{N(0,1)}$. Extracting an i-vector from the total-variability subspace is essentially a \textit{maximum a-posteriori adaptation}~(MAP) of $\textbf{w}$ in the subspace defined by $\textbf{T}$. An efficient procedure for the optimization of the total-variability subspace $\textbf{T}$ and subsequent extraction of i-vectors is described Dehak~\emph{et al.}~\cite{Dehak2010,Kenny2008}. In this paper, the pooled total-variability approach is used for i-vector feature extraction where the total-variability subspace~(${R_{w}}^{telmic} = 500$) is trained on telephone and microphone speech utterances together.
\begin{figure*}[ht]	
\centering
\scalebox{1.0}{
\includegraphics[width=14cm, height=7cm]{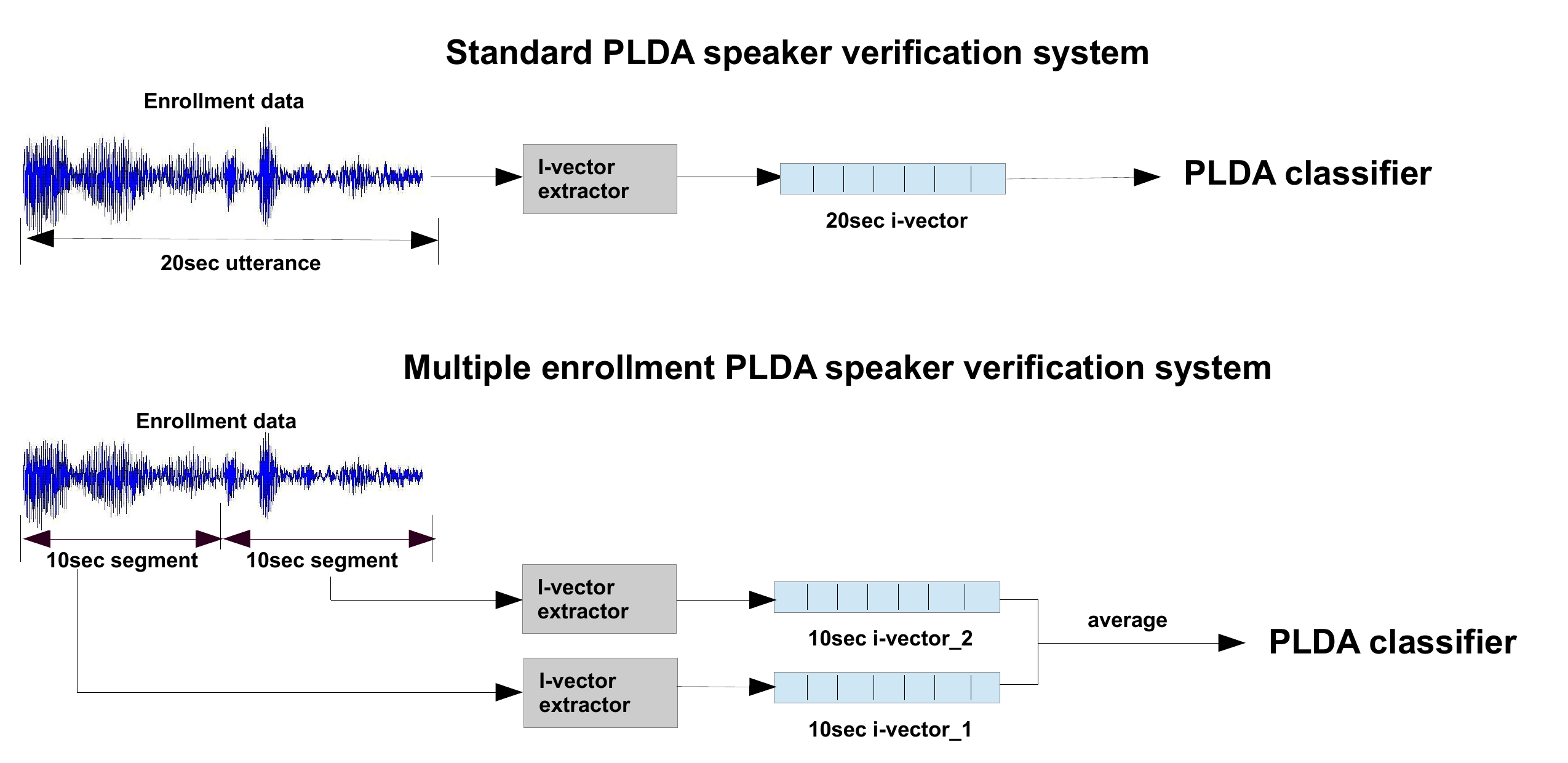}}
\caption{\emph{A block diagram of utterance-partitioning-based PLDA speaker verification.}}
\label{fig:multiple enrolment approach}
\end{figure*}
\section{Short utterance variance added i-vector features} \label{sec:SUV added i-vectors}
The long-length utterance i-vectors have speaker and session variations whereas short-length i-vectors have speaker, session and a lot of utterance variations. Thus, during development for SUV-PLDA, utterance variance is captured using the inner product of the difference between the full- and short-length i-vectors, and it is artificially added to full-length utterances and the simulated SUV is modelled using the PLDA approach. The short utterance variance matrix, $\textbf{S}_{SUV}$, can be calculated as follows,
\begin{eqnarray}
    \textbf{S}_{SUV} = \frac{1}{N}\sum_{n=1}^{N}\textbf{A}^{T}(\textbf{w}^{full}_{n} - \textbf{w}^{short}_{n})\textbf{A}^{T}(\textbf{w}^{full}_{n} - \textbf{w}^{short}_{n})^{T} \label{eqn:S_SUV estimation}
\end{eqnarray}
where the estimation of the LDA matrix, $\textbf{A}$ is detailed in our previous work~\cite{Kanagasundaram2014(Submitted)}. For $\textbf{S}_{SUV}$ estimation, the actual definition of what constitutes a full and/or short-length utterance needs to be established. For this research, we have defined full-length to be NIST standard utterance length, and in order to capture the SUV, short utterance length was selected as 20 sec. The SUV decorrelated matrix, $\textbf{D}$, is calculated using the Cholesky decomposition of $\textbf{D}\textbf{D}^{T} = \textbf{S}_{SUV}$. A random vector with utterance variation information can be generated if random normally independently distributed vector, $\textbf{d}$, with $\mu=0.0$ and $\sigma=1.0$ is multiplied by the SUV decorrelated matrix, $\textbf{D}$. The SUV-added full-length development vectors can be estimated as follows,
\begin{eqnarray}
    \textbf{w} = \textbf{w}_{full} + {\textbf{D}}^{T}\textbf{d} \label{eqn:SUV only addition}
\end{eqnarray}
After the SUV-added full-length i-vectors are extracted, length-normalized GPLDA model parameters are estimated in as described in Section~\ref{sec:len-norm GPLDA}.
\section{Length-normalized GPLDA system} \label{sec:len-norm GPLDA}
\subsection{PLDA modelling}
In this paper, we have chosen length-normalized GPLDA, as it is also a simplified and computationally efficient approach~\cite{Garcia-Romero2011}. The length-normalization approach is detailed by Garcia-Romero~\emph{et al.}~\cite{Garcia-Romero2011}, and this approach is applied on development and evaluation data prior to GPLDA modelling. A speaker and channel dependent length-normalized i-vector, $\hat{\textbf{w}}_{r}$ can be defined as,
\begin{eqnarray} \label{PLDA model}
    \hat{\textbf{w}}_{r} & = & \bar{\hat{\textbf{w}}} + \textbf{U}_{1}\textbf{x}_{1} + \boldsymbol{\varepsilon}_{r}
\end{eqnarray}
where for given speaker recordings $r = 1,.....R$; $\textbf{U}_{1}$ is the eigenvoice matrix, $\textbf{x}_1$ is the speaker factors and $\boldsymbol{\varepsilon}_{r}$ is the residuals. In the PLDA modelling, the speaker specific part can be represented as $\bar{\textbf{w}} + \textbf{U}_{1}\textbf{x}_{1}$, which represents the between speaker variability. The covariance matrix of the speaker part is $\textbf{U}_{1}{\textbf{U}_{1}}^{T}$. The channel specific part is represented as $\boldsymbol{\varepsilon}_{r}$, which describes the within speaker variability. The covariance matrix of channel part is ${\boldsymbol{\Lambda}}^{-1}$. We assume that precision matrix~($\boldsymbol{\Lambda}$) is full rank. Prior to GPLDA modelling, standard LDA approach is applied to compensate the additional channel variations as well as reduce the computational time~\cite{Kanagasundaram2012a}.
\subsection{GPLDA scoring} \label{sec:PLDA scoring}
Scoring in GPLDA speaker verification systems is conducted using the batch likelihood ratio between a target and test i-vector~\cite{Kenny2010}. Given two i-vectors, $\textbf{w}_{target}$ and $\textbf{w}_{test}$, the batch likelihood ratio can be calculated as follows,
\begin{eqnarray}
    \ln\frac{P(\textbf{w}_{target},\textbf{w}_{test}\mid H_{1})}{P(\textbf{w}_{target}\mid H_{0})P(\textbf{w}_{test}\mid H_{0})}
\end{eqnarray}
where $H_{1}$ denotes the hypothesis that the i-vectors represent the same speakers and $H_{0}$ denotes the hypothesis that they do not.
\section{Utterance-partitioning-based PLDA speaker verification} \label{sec: utternace parition method}
A single long utterance is commonly used as enrolment data in PLDA speaker verification system. It was previously found that short utterance i-vectors have speaker, session and utterance variations. It is hypothesised that if long-duration speech data is partitioned into short utterances and if average of short utterance i-vectors is estimated, this approach would compensate the utterance variations. In this paper, in order to test this hypothesis, long enrolled-utterance are partitioned into short utterances and multiple i-vectors are extracted and average of extracted i-vectors is used as enrolled i-vector. A block diagram of utterance-partitioning-based PLDA speaker verification is shown in Figure~\ref{fig:multiple enrolment approach}. Though truncated short utterances are extracted from same speaker and session, every short utterance i-vectors have different behaviour due to linguistic content and the averaging of multiple short utterance i-vector can be used compensate the linguistic content variation.
\begin{table}
\caption{\label{tab:LDA multiple enrollment PLDA speaker verification}\emph{Comparison of standard GPLDA and utterance-partitioning-based GPLDA speaker verification systems on the common set of the 2008 and 2010 NIST SRE truncated conditions. (a) NIST 2008 short2-short3 truncated condition (b) NIST 2010 core-core truncated condition. The best performing systems by both EER and DCF are highlighted across each row.}}
\begin{center}
\subfigure[\emph{NIST 2008 short2-short3 truncated condition}]{
\scalebox{1.0}{
\begin{tabular}{l c c} \hline
  \textbf{Evaluation} & \multirow{2}{*}{\textbf{EER}} & \multirow{2}{*}{\textbf{DCF}} \\
  \textbf{utterance lengths} & &  \\ \hline
\multicolumn{3}{l}{\textbf{Standard GPLDA system (Baseline)}} \\
  10sec-10sec & 15.90\% & 0.0656 \\
  20sec-10sec & 13.26\% & \textbf{0.0549} \\ \hline
\multicolumn{3}{l}{\textbf{Utterance-partitioning-based GPLDA system}} \\
  10sec (2)-10sec & \textbf{12.60\%} & 0.0552 \\ \hline
\end{tabular} }}
\label{tbl:A}
\end{center}

\begin{center}
\subfigure[\emph{NIST 2010 core-core truncated condition}]{
\scalebox{1.0}{
\begin{tabular}{c c c} \hline
  \textbf{Evaluation}& \multirow{2}{*}{\textbf{EER}} & \multirow{2}{*}{\textbf{DCF}} \\
  \textbf{utterance lengths} &  &  \\ \hline
\multicolumn{3}{l}{\textbf{Standard GPLDA system (Baseline)}} \\
  10sec-10sec& 15.98\% & 0.0693 \\
  20sec-10sec & 13.85\% & 0.0608 \\ \hline
\multicolumn{3}{l}{\textbf{Utterance-partitioning-based GPLDA system}} \\
  10sec (2)-10sec & \textbf{13.01\%} & \textbf{0.0588} \\ \hline
\end{tabular} }}
\label{tbl:B}
\end{center}
\end{table}
\section{Experimental methodology} \label{sec:method}
The proposed methods were evaluated using the the NIST 2008 and NIST 2010 SRE corpora. The shortened evaluation utterances were obtained by truncating the NIST 2008 \emph{short2}-\emph{short3} and NIST 2010 \emph{core}-\emph{core} conditions to the specified length of active speech for both enrolment and verification. Prior to truncation, the first 20 seconds of active speech were removed from all utterances to avoid capturing similar data across multiple utterances. For NIST 2008, the performance was evaluated using the equal error rate (EER) and the minimum decision cost function (DCF), calculated using $\emph{C}_{miss} = 10$, $\emph{C}_{FA} = 1$, and $\emph{P}_{target} = 0.01$~\cite{NIST2008}. The performance for the NIST 2010 SRE was evaluated using the EER and the old minimum DCF~(${DCF}_{old}$), calculated using $\emph{C}_{miss} = 10$, $\emph{C}_{FA} = 1$, and $\emph{P}_{target} = 0.01$, where evaluation was performed using the \emph{telephone}-\emph{telephone} condition~\cite{NIST2010}.

We have used 13 feature-warped MFCC with appended delta coefficients and two gender-dependent UBMs containing 512 Gaussian mixtures throughout our experiments. The UBMs were trained on telephone and microphone speech from NIST 2004, 2005, and 2006 SRE corpora, and then used to calculate the Baum-Welch statistics before training a gender dependent total-variability subspace of dimension $R_{w} = 400$. The pooled total-variability representation and the GPLDA parameters were trained using telephone and microphone speech data from NIST 2004, 2005 and 2006 SRE corpora as well as Switchboard II which includes 1386 female and 1117 male speakers. We empirically selected the number of eigenvoices~($N_{1}$) equal to 120 as best value according to speaker verification performance over an evaluation set. 150 eigenvectors were selected for LDA estimation. S-normalisation was applied for experiments, and randomly selected telephone and microphone utterances from NIST 2004, 2005 and 2006 were pooled to form the S-normalisation dataset~\cite{Shum2010}.
\begin{table}
\caption{\label{tab:LDA multiple enrollment SUV PLDA speaker verification}\emph{Comparison of SUV-GPLDA and utterance-partitioning-based SUV-GPLDA speaker verification systems on the common set of the 2008 and 2010 NIST SRE truncated conditions. (a) NIST 2008 short2-short3 truncated condition (b) NIST 2010 core-core truncated condition. The best performing systems by both EER and DCF are highlighted across each row.}}
\begin{center}
\subfigure[\emph{NIST 2008 short2-short3 truncated condition}]{
\scalebox{1.0}{
\begin{tabular}{l c c} \hline
  \textbf{Evaluation}& \multirow{2}{*}{\textbf{EER}} & \multirow{2}{*}{\textbf{DCF}} \\
  \textbf{utterance lengths} &  &  \\ \hline
\multicolumn{3}{l}{\textbf{Standard SUV-GPLDA system}} \\
  10sec-10sec & 14.58\% & 0.0624 \\
  20sec-10sec & 12.35\% & 0.0523 \\ \hline
\multicolumn{3}{l}{\textbf{Utterance-partitioning-based SUV-GPLDA system}} \\
  10sec (2)-10sec & \textbf{12.05\%} & \textbf{0.0519} \\ \hline
\end{tabular} }}
\label{tbl:A}
\end{center}

\begin{center}
\subfigure[\emph{NIST 2010 core-core truncated condition}]{
\scalebox{1.0}{
\begin{tabular}{l c c} \hline
  \textbf{Evaluation}& \multirow{2}{*}{\textbf{EER}} & \multirow{2}{*}{\textbf{DCF}} \\
  \textbf{utterance lengths} &  &  \\ \hline
\multicolumn{3}{l}{\textbf{Standard SUV-GPLDA system}} \\
  10sec-10sec & 14.70\% & 0.0672 \\
  20sec-10sec & 12.00\% & 0.0578 \\ \hline
\multicolumn{3}{l}{\textbf{Utterance-partitioning-based SUV-GPLDA system}} \\
  10sec (2)-10sec & \textbf{11.58\%} & \textbf{0.0555} \\ \hline
\end{tabular} }}
\label{tbl:B}
\end{center}
\end{table}
\section{Results and discussions} \label{sec:results and discussions}
\subsection{Utterance-partitioning-based GPLDA system} \label{sec:multiple enrolment LDA-projected system}
In this section, the performance of standard LDA-projected PLDA and utterance-partitioning-based LDA-projected GPLDA were compared on NIST 2008 and 2010 truncated conditions. Standard LDA-projected GPLDA was evaluated on 10sec-10sec and 20sec-10sec conditions. For utterance-partitioning-based LDA-projected GPLDA system, 20sec enrolment utterance was truncated into two 10sec utterances and average of 10sec i-vectors was used as enrolment i-vector. Table~\ref{tab:LDA multiple enrollment PLDA speaker verification} compares the performance of utterance-partitioning-based LDA-projected GPLDA system against standard LDA-projected GPLDA system. Utterance-partitioning-based LDA-projected system shows improvement over standard LDA-projected system. Based upon these results, it is believe that though truncated short enrolled-utterances are extracted from same speaker and session, every short enrolled-utterance i-vectors have different behaviour due to linguistic content and the averaging of multiple short enrolled-utterance i-vectors can be used compensate the linguistic content variation.
\subsection{Utterance-partitioning-based SUV-GPLDA system} \label{sec:multiple enrolment LDA-projected system}
In our previous studies, we have found that SUV-added GPLDA approach can effectively model the short utterance variance~\cite{Kanagasundaram2014(Submitted)}. In this section, short utterance variance was studied with utterance-partitioning-based GPLDA system. Table~\ref{tab:LDA multiple enrollment SUV PLDA speaker verification} compares the performance of utterance-partitioning-based LDA-projected SUV-GPLDA system against LDA-projected SUV-GPLDA system. It can be clearly seen that utterance-partitioning-based LDA-projected SUV-GPLDA system shows improvement over LDA-projected SUV-GPLDA system. When utterance-partitioning-based LDA-projected SUV-GPLDA system is compared against standard LDA-projected GPLDA system from Table~\ref{tab:LDA multiple enrollment PLDA speaker verification} and~\ref{tab:LDA multiple enrollment SUV PLDA speaker verification}, utterance-partitioning-based LDA-projected SUV-GPLDA system shows relative improvement of 9\% and 16\% in EER for NIST 2008 and NIST 2010 truncated 10sec-10sec evaluation condition.
\section{Conclusion}~\label{sec:conclusion}
This paper studied the PLDA speaker verification approach with utterance partitioning and SUV modelling approaches. Our experimental studies have found that instead of using single long-utterance as enrolment data, if long enrolled-utterance was partitioned into multiple short utterances and average of short utterance i-vectors was used as enrolled data, that improved the GPLDA speaker verification. This is because short utterance i-vectors have speaker, session and utterance variations, and utterance-partitioning approach compensates the utterance variation. SUV-GPLDA speaker was also studied with utterance-partitioning approach, and utterance-partitioning-based LDA-projected SUV-GPLDA system showed relative improvement of 9\% and 16\% in EER for NIST 2008 and NIST 2010 truncated 10sec-10sec evaluation condition as utterance-partitioning approach compensates the utterance variation and SUV modelling approach compensates the mismatch between full-length development data and short-length evaluation data.

\section{Acknowledgements}
This project was supported by an Australian Research Council (ARC) Linkage grant LP130100110.
%
\eightpt
\bibliographystyle{IEEEtran}
\bibliography{research}

\begin{thebibliography}{10}
\providecommand{\url}[1]{#1}
\csname url@samestyle\endcsname
\providecommand{\newblock}{\relax}
\providecommand{\bibinfo}[2]{#2}
\providecommand{\BIBentrySTDinterwordspacing}{\spaceskip=0pt\relax}
\providecommand{\BIBentryALTinterwordstretchfactor}{4}
\providecommand{\BIBentryALTinterwordspacing}{\spaceskip=\fontdimen2\font plus
\BIBentryALTinterwordstretchfactor\fontdimen3\font minus
  \fontdimen4\font\relax}
\providecommand{\BIBforeignlanguage}[2]{{%
\expandafter\ifx\csname l@#1\endcsname\relax
\typeout{** WARNING: IEEEtran.bst: No hyphenation pattern has been}%
\typeout{** loaded for the language `#1'. Using the pattern for}%
\typeout{** the default language instead.}%
\else
\language=\csname l@#1\endcsname
\fi
#2}}
\providecommand{\BIBdecl}{\relax}
\BIBdecl

\bibitem{Vogt2008a}
R.~Vogt, B.~Baker, and S.~Sridharan, ``Factor analysis subspace estimation for
  speaker verification with short utterances,'' in \emph{Interspeech 2008},
  Brisbane, Australia, September 2008.

\bibitem{Kanagasundaram2011}
A.~Kanagasundaram, R.~Vogt, B.~Dean, S.~Sridharan, and M.~Mason, ``i-vector
  based speaker recognition on short utterances,'' in \emph{Proceed. of
  INTERSPEECH}.\hskip 1em plus 0.5em minus 0.4em\relax International Speech
  Communication Association (ISCA), 2011, pp. 2341--2344.

\bibitem{McLaren2010b}
M.~McLaren, R.~Vogt, B.~Baker, and S.~Sridharan, ``Experiments in {SVM}-based
  speaker verification using short utterances,'' in \emph{Proc. Odyssey
  Workshop}, 2010, pp. 83--90.

\bibitem{Kanagasundaram2012a}
A.~Kanagasundaram, R.~Vogt, D.~Dean, and S.~Sridharan, ``{PLDA} based speaker
  recognition on short utterances,'' in \emph{The Speaker and Language
  Recognition Workshop (Odyssey 2012)}.\hskip 1em plus 0.5em minus 0.4em\relax
  ISCA, 2012.

\bibitem{Kanagasundaram2014(Submitted)}
A.~Kanagasundaram, D.~Dean, and S.~Sridharan, ``Improving {PLDA} speaker
  verification with limited development data,'' in \emph{IEEE Int. Conf. on
  Acoustics, Speech and Signal Processing}, 2014.

\bibitem{Kanagasundaram2013b}
A.~Kanagasundaram, D.~Dean, J.~Gonzalez-Dominguez, S.~Sridharan, D.~Ramos, and
  J.~Gonzalez-Rodriguez, ``Improving short utterance based i-vector speaker
  recognition using source and utterance-duration normalization techniques,''
  in \emph{Proceed. of INTERSPEECH}.\hskip 1em plus 0.5em minus 0.4em\relax
  International Speech Communication Association (ISCA), 2013.

\bibitem{Dehak2010}
N.~Dehak, P.~Kenny, R.~Dehak, P.~Dumouchel, and P.~Ouellet, ``Front-end factor
  analysis for speaker verification,'' \emph{Audio, Speech, and Language
  Processing, IEEE Transactions on}, vol.~PP, no.~99, pp. 1 --1, 2010.

\bibitem{Kenny2013}
P.~Kenny, T.~Stafylakis, P.~Ouellet, M.~Alam, and P.~Dumouchel, ``{PLDA} for
  speaker verification with utterances of arbitrary duration,'' in \emph{IEEE
  Int. Conf. on Acoustics, Speech and Signal Processing}, 2013.

\bibitem{Kanagasundaram2013(tobesubmitted)}
A.~Kanagasundaram, D.~Dean, S.~Sridharan, J.~Gonzalez-Dominguez, D.~Ramos, and
  J.~Gonzalez-Rodriguez, ``Improving short utterance i-vector speaker
  recognition using utterance variance modelling and compensation techniques,''
  in \emph{Speech Communication}.\hskip 1em plus 0.5em minus 0.4em\relax
  Publication of the European Association for Signal Processing (EURASIP),
  2014.

\bibitem{Dehak2009a}
N.~Dehak, R.~Dehak, P.~Kenny, N.~Brummer, P.~Ouellet, and P.~Dumouchel,
  ``Support vector machines versus fast scoring in the low-dimensional total
  variability space for speaker verification,'' in \emph{Proceedings of
  Interspeech}, 2009, p. 1559~1562.

\bibitem{Kenny2008}
P.~Kenny, P.~Ouellet, N.~Dehak, V.~Gupta, and P.~Dumouchel, ``A study of
  inter-speaker variability in speaker verification,'' \emph{IEEE Transactions
  on Audio, Speech, and Language Processing}, vol.~16, no.~5, pp. 980--988,
  2008.

\bibitem{Garcia-Romero2011}
D.~Garcia-Romero and C.~Espy-Wilson, ``Analysis of i-vector length
  normalization in speaker recognition systems,'' in \emph{International
  Conference on Speech Communication and Technology}, 2011, pp. 249--252.

\bibitem{Kenny2010}
P.~Kenny, ``Bayesian speaker verification with heavy tailed priors,'' in
  \emph{Proc. Odyssey Speaker and Language Recogntion Workshop, Brno, Czech
  Republic}, 2010.

\bibitem{NIST2008}
\BIBentryALTinterwordspacing
``The {NIST} year 2008 speaker recognition evaluation plan,'' NIST, Tech. Rep.,
  2008. [Online]. Available:
  \url{http://www.itl.nist.gov/iad/mig/tests/sre/2008/}
\BIBentrySTDinterwordspacing

\bibitem{NIST2010}
\BIBentryALTinterwordspacing
``The {NIST} year 2010 speaker recognition evaluation plan,'' NIST, Tech. Rep.,
  2010. [Online]. Available:
  \url{www.itl.nist.gov/iad/mig/tests/sre/2010/index.html}
\BIBentrySTDinterwordspacing

\bibitem{Shum2010}
S.~Shum, N.~Dehak, R.~Dehak, and J.~Glass, ``Unsupervised speaker adaptation
  based on the cosine similarity for text-independent speaker verification,''
  \emph{Proc. Odyssey}, 2010.

\end{thebibliography}
\end{document}